\documentclass[preprint,12pt]{elsarticle}

\usepackage{amssymb}
\usepackage{mathtools}
\usepackage{siunitx}
\newcommand{\pace}{{\textsc{Pace3D}}} 
\usepackage[doublespacing]{setspace}

\journal{Scripta Materialia}

\begin{document}
\begin{frontmatter}



\title{A U-Net-Based Self-Stitching Method for Generating Periodic Grain Structures}


\author[1,2,3,4]{Ye Ji\corref{cor1}}
\ead{silentmlight@gmail.com}
\author[4]{Arnd Koeppe}
\author[4,5]{Patrick Altschuh}
\author[4]{Lars Griem}
\author[4]{Deepalaxmi Rajagopal}
\author[4,5]{Britta Nestler}
\author[1,2,3]{Weijin Chen}
\author[1,2,3]{Yi Zhang}
\author[1,2,3]{Yue Zheng\corref{cor1}\fnref{fn1}}
\ead{zhengy35@mail.sysu.edu.cn}

\cortext[cor1]{Corresponding author}
\fntext[fn1]{Permanent address: State Key Laboratory of Optoelectronic Materials and Technologies, School of Physics, Sun Yat-sen University, Guangzhou, 510275, Guangdong, China.}

\affiliation[1]{organization={Centre for Physical mechanics and Biophysics, School of Physics, Sun Yat-sen University},
            addressline={Xin'gang West Road 135}, 
            city={Guangzhou},
            postcode={510275}, 
            state={Guangdong},
            country={China}}

\affiliation[2]{organization={Guangdong Provincial Key Laboratory of Magnetoelectric Physics and Devices, School of Physics},
            addressline={Xin'gang West Road 135}, 
            city={Guangzhou},
            postcode={510275}, 
            state={Guangdong},
            country={China}}

\affiliation[3]{organization={State Key Laboratory of Optoelectronic Materials and Technologies, School of Physics},
            addressline={Xin'gang West Road 135}, 
            city={Guangzhou},
            postcode={510275}, 
            state={Guangdong},
            country={China}}

\affiliation[4]{organization={Institute for Applied Materials–Microstructure Modelling and Simulation, Karlsruhe Institute of Technology},
            addressline={Strasse am Forum 7}, 
            city={Karlsruhe},
            postcode={76131}, 
            state={Baden-Württemberg},
            country={Germany}}

\affiliation[5]{organization={Institute for Digital Materials Science, Karlsruhe University of Applied Sciences},
            addressline={Moltkestr. 30}, 
            city={Karlsruhe},
            postcode={76133}, 
            state={Baden-Württemberg},
            country={Germany}}
            
\begin{abstract}
When modeling microstructures, the computational resource requirements increase rapidly as the simulation domain becomes larger. As a result, simulating a small representative fraction under periodic boundary conditions is often a necessary simplification. However, the truncated structures leave nonphysical boundaries, which are detrimental to numerical modeling. Here, we propose a self-stitching algorithm for generating periodic structures, demonstrated in a grain structure. The main idea of our algorithm is to artificially add structural information between mismatched boundary pairs, using the hierarchical spatial predictions of the U-Net. The algorithm provides an automatic and unbiased way to obtain periodic boundaries in grain structures and can be applied to porous microstructures in a similar way.
\end{abstract}


\end{frontmatter}

Computational simulation, an economical and powerful research paradigm, plays an important role in explaining experiments, gaining insights into fundamental physical mechanisms towards designing materials and components. In materials science, one of the essential tasks for computational simulation is to reveal the structure-property linkage through homogenization approaches considering that full resolution models are extremely expensive to compute. In this regard, setting a representative geometry that is large enough to reflect the macroscopic properties is the first step. A larger simulation domain usually contains more information about the system, but at the same time involves higher computational costs that, in a power law \cite{caloComputationalComplexityMemory2011}, are scaled with the size of the computational domain, which rapidly becomes unaffordable again. Therefore, a representative volume element with periodic boundary conditions is a practical solution \cite{pivovarovPeriodicBoundaryConditions2019}. However, real microstructures do not always ensure periodicity, except for few cases, such as crystal structures. Each representative volume element is part of a larger system, which truncates the boundaries. This leads to an inconsistency between the corresponding boundaries and produces additional nonphysical effects. For example, the Pores Per Inch (PPI) of a metal foam is a widely used parameter in industrial applications to characterize porous materials. However, the pores on the boundaries are ignored when determining the PPI \cite{jamshidi3DComputationalMethod2023}. 
Therefore, an approach that allows for the identification of boundary pores in a periodic structure will result in a more accurate PPI calculation.

There are many existing methods to generate periodic structures from scratch \cite{richterMote3DOpensourceToolbox2017}. However, the generation of periodic boundaries comparable with reference structures is less studied. Mathematically, converting a non-periodic structure into a periodic structure is equivalent to interpolating the mismatching boundaries by a zone that preserves the intrinsic characteristics of the larger system. 
The desired interpolation should go beyond the direct truncation, the air layer \cite{rutterChargedSurfacesSlabs2021}, and the polynomial interpolation \cite{nguyenImposingPeriodicBoundary2012,wangNovelApproachImpose2017}. Neural networks are successful and powerful tools that meet such requirements. The neural network is attracting considerable attention, as it has been shown to be a universal mapping tool that approximates any function \cite{zhouUniversalityDeepConvolutional2020}. This ability allows the extraction of the characteristic features of data, as well as the generation of a new structure based on the combination of learned features \cite{kalidindiHandbookBigData2020}. Here, the interpolation between two boundaries can be seen as a problem of modifying existing structures. Therefore, it is promising to apply data-driven solution schemes to the current problem.

In this work, we propose a U-Net based self-stitching algorithm to generate a periodic grain structure from a non-periodic grain structure. As shown in Fig. 1, the algorithm consists of a training phase and a prediction phase. In the training phase, the U-Net is forced to reconstruct the masked structure, while in the prediction phase, the boundaries to be joined are first arranged in an individual structure and then connected by U-Net. A periodic version of the corresponding non-periodic structure can be obtained by a further step of simple manipulation. The proposed method is examined in one representative direction of a 2D grain structure. The U-Net is set up and trained, and its performance is evaluated. We emphasize that the mask width is a crucial parameter in the algorithm. Our results prove that the proposed algorithm is able to generate periodic boundaries in grain samples and paves the way for similar problems in a wide variety of systems.

The multi-phase field method \cite{nestlerPhasefieldModelMultiphase2008} can be used to compute the evolution of polycrystalline grain structures with a large amount of grains with different crystallographic orientations, the theory of which is briefly provided in the supplementary data. The approach works in 2D and 3D and is capable to describe normal and abnormal grain growth for isotropic as well as anisotropic interface conditions and takes into account various physical influences such as heat treatment, mechanical load or diffusion. Grain growth processes are performed with the simulation framework {\pace}~ \cite{hotzerParallelMultiphysicsPhasefield2018}, which is developed and maintained in the research group of the IDM at the Karlsruhe University of Applied Sciences. The parameters incorporated in the simulation can be found in \cite{janssensComputationalMaterialsEngineering2007}. A simulation domain of 4000 $\times$ 4000 numerical cells is set up and 60000 randomly distributed grains are initialized by a Voronoi-tessellation algorithm in a preprocessing step. After the simulation, 20000 small patches are sampled from an intermediate time step. These grain samples are skeletonized for further data analysis applications. Results of phase-field simulations of grain structures serve as reference data sets for the developed U-Net approach.

 U-Nets were initially developed for classification problem, but with slight modification, it could also solve our spatial regression problem. The details of U-Net are shown in Fig. S1 in the suppplementary materials. In the proposed self-stitching method, the U-Net should be able to generate a reasonable structure between any two given boundaries. This ability will be acquired through training with artificially masked data. If a grain sample is intersected with two non-intersecting lines, the right boundary of the left part and the left boundary of the right part do not coincide, but they can always be connected by the inner part. In this investigation, the two boundaries are connected by the inserted structure. Based on this straightforward idea, we can manually mask the inner part and force the U-Net to reconstruct the masked structure on the basis of the unmasked structure. In principle, the shape of the boundary is not limited, but straight boundaries occur naturally in computational modeling. In addition, with a two-dimensional grain sample, it is convenient to use a rectangular mask. The long edges of the mask area are considered as the newly created boundaries. That means that the self-stitching is done only in the direction that is parallel to the short edge. We believe that once self-stitching is achieved in one direction, periodicity in two-directions could be achieved by following the same procedure in each direction. In addition, the mask is placed in the middle of the grain sample, for reasons of information symmetry. 

As shown in Fig. 2a, the black stripe means that all values are set to zero, which acts as the mask. The width of the mask is 16 cells in the horizontal direction and throughout the vertical direction. The masked sample is used as input to U-Net, and the original sample is the target for the output, as shown in Fig. 2b. The masked area occupies a small part of the whole structure, ensuring enough structural information the rest area possesses is passed to the U-Net. The dataset contains 20000 pairs of masked samples and their original counterparts. The dataset is divided into the training dataset, the validation dataset, and the test dataset. The split ratio is 0.6:0.2:0.2. Since data are categorical, we use sparse categorical cross entropy as the loss function. A batch size of 8 is chosen along with the default Adam optimizer to train U-Net for 10 epochs on two GeForce GTX 1080 GPU nodes. After the training, the prediction for the masked area in Fig. 2a is shown in Fig. 2c. Since the prediction is close to the ground truth, the difference between the two is shown in Fig. 2d for visual clarity. 
In Fig. 2d, the colored pixels occur only within the masked area, which means that the rest of the structure is completely preserved. 
Here, the red pixel (true negative) means that this point should belong to a boundary, but the prediction varies from the ground truth; the blue pixel (false positive) refers to the fact that this point should not belong to a boundary, but the prediction is right. Despite the presence of falsely predicted values, it is observed that the red and blue pixels are usually correlated, which could be translated as the predicted boundaries are shifted. The boundary shift is tolerable, compared to randomness. Fig. 2e shows the loss curve, where both the training loss and the validation loss  decrease. The training loss is slightly less than the validation loss, which indicates that any overfitting is not severe. A modified spatial accuracy is also calculated and displaed in Fig. 2f based on the number of correctly predicted points versus total points within the masked area. Fig. 2f shows that the trend in accuracy is consistent with the loss curve, and  that the overall accuracy is over 90\%, despite the boundary shifts. Therefore, it can be assumed that U-Net correctly scans the existing grain structures to infer and reproduce the missing structure from the given structural information. 

With this generative capability, we then investigate its feasibility in producing periodic boundaries. The process starts from a normal grain sample, illustrated in Fig. 3a. The left half of the sample is indicated by a green box, while the right half of the sample is indicated by a blue box. The red dotted line marks the borders of the mask and is helpful to visually follow the manipulation. In Fig. 3b, the left and right parts are swapped so that the boundary pair to be stitched is joined. Near the central line, the mismatch is visible. Then, a masked area is inserted in Fig. 3c, which increases the total width of the sample. The width of the mask is identical to the one used in the training process. The sample is then cropped along the red lines on both sides. The remaining structure is shown in Fig. 3d. The cropped size matches the shape of the input data of the U-Net, so that the information can be predicted within the mask, as shown in Fig. 3e. Finally, the structure previously cut out can be restored. In Fig. 3f, we see a periodic sample in the horizontal direction. The imperfection is that some grain boundaries are not fully closed in the stitched sample. However, the predicted structure outlines the possible structure where the line segment can be extended a little to close most of the grains. This issue will be discussed in the next section. It is worth noting that in the masked area, not only the straight lines are connected, but also triple points are added, which demonstrates its power. 

Since there is no ground truth that can be compared to the stitched structure, it is not easy to quantify the performance. On the one hand, the generated area should be somehow similar to the rest of the original sample, by definition. This is the desired result. On the other hand, the stitched structure should be created other than by just putting the edges together, without interpolation. This is the worst result, which should be surpassed by the method. Therefore, it is helpful to compare the original sample, the switched sample, and the restored sample. Using the results from Figs. 3a, 3b, and 3f, we perform a watershed segmentation and show the results in Figs. 4a, 4b, and 4c, respectively. The watershed algorithm \cite{gostickOpenPNMPoreNetwork2016} defines the regions by recursively searching around the peaks in an elevation map, which could be obtained from a binary image, by a Euclidean distance transformation. Watershed segmentation captures the local feature of a structure, ensuring that the segmented structure is closed. In this way, the evaluation of boundaries is transformed into a comparison of the regularity of the shape of the grains. The segmentation results are not spatially comparable point-to-point, but it can be seen that only in Fig. 4b, as highlighted in the magnification box, boundaries with sharp spikes are formed. These spikes are not seen in the original sample, so they are not desirable for a realistic grain sample. 

To measure the difference, we introduce the local curvature \cite{kremeyerCellularAutomataInvestigations1998}, which is defined by counting the number of phases in  the neighborhood of a point. The grain boundary maps illustrating the curvature distributions of Figs. 4a, 4b, and 4c are given in Figs. 4d, 4e, and 4f, respectively. We note that the curvature distributions must be calculated based on the segmented map, as this is the only way to distinguish the grains. It can be observed that only the boundaries are non-zero, and the brightness of junctions is higher than that of the straight boundaries. These features are useful for detecting the irregular boundary shown magnified in the yellow box of Fig. 4e. In Fig. 4f, no irregular boundary is found, which confirms that the U-Net-assisted stitching gives better results than the direct concatenation of two boundaries. 

So far, we have shown that the U-Net can fill the mask area with proper training. The question arises whether the U-Net is hyperparameter sensitive. In other words, is there a requirement that ensures that U-Net provides meaningful predictions? During the preprocessing step, the width of the mask seems to be an essential parameter because the convolutional neural network always captures the local feature, which implies that there should be an effective action length. The valid range of width is between zero and positive infinity. If it is small, there is not enough space for the connecting line segment, resulting in a distorted configuration. In the extreme case, there is no space, which was discussed in Figs. 4b and 4e. If, on the other hand, it is so large that exceeds the width of the entire sample, the expected information is even more than the given information, which is unlikely. The extreme case of an infinite width, corresponding to generating an RVE from scratch, is also impractical. 

In the previous results, the width of the mask is 16- cells. A narrow width is tested for comparison, and the corresponding results are plotted in Fig. 5a-c. A wide width is tested, and the related results are plotted in Fig. 5d-f. Only the most representative configurations of the workflow are shown in these figures. In this context, Fig. 5a and 5d are the counterparts of Fig. 2c; Fig. 5b and 5e are the counterparts of Fig. 3f, and Fig. 5c and 5f are the counterparts of Fig. 4f. The same visualization styles from the previous section are retained. In addition, the shape of the mask is explicitly marked. As expected, the boundary is connected in the narrow case, while it is not connected in the wide case, although the lines extend into the masked area. It is noteworthy that the boundaries in Fig. 5e overlap with the masked area several times. In Fig. 5f, however, they shrink. It is likely that watershed segmentation tends to eliminate incomplete boundaries here. Recalling that the incomplete boundaries are closed by the watershed operation in the previous case, the two situations suggest that the mask width should not be too large. In addition, the training loss, accuracy, and accumulative curvature from the three cases are summarized and shown in Figs. 6g-i. It can be observed that the loss increase monotonically with the mask width, while the accuracy decreases in a similar manner. In Fig. 5i, the green line represents the average accumulative curvature in the masked area, while the blue dashed line denotes the reference value in the unmasked area. It can be clearly seen that the value of the generated structure intersects with the reference value, suggesting that the width should be neither too narrow nor too wide. Around the intersecting point, a smaller mask width outperforms a larger one because the incomplete boundaries tend to close rather than disappear after the watershed operation.

In this paper, we proposed a U-Net-based algorithm to convert a nonperiodic microstructure into a periodic microstructure. The validity of this method is verified using 2D grain samples along one direction. The U-Net is trained with masked data as input and the original data as ground truth. The training converges and successfully predicts the masked structure. A series of manipulations is performed to obtain a variant structure containing two nonmatching boundaries, separated by a masked area. The masked area in the variant structure is filled, and a periodic structure is obtained by adding the cropped parts back. A metric for evaluating the quality of the generated pattern is developed and the effect of the mask width, based on this metric, is studied. It is assumed that there is an optimal value of the width of the mask.

The method uses a neural network to detect and stitch the boundaries. It is not limited to the grain structure but can be used for many porous structures, and even more structures with statistical homogeneity. It should also be able to extend to 3D systems for practical application. In any case, the mask width should be essential, which may rely on the characteristic length of the dominant features. Further, for a microstructure where multiscale features exist, systematic work will be needed in order to investigate what determines the optimum mask width.

\section*{Acknowledgements}
We thank the 2020 'Helmholtz-OCPC Program for the involvement of postdocs in bilateral collaboration projects' for their financial support that enabled this study. This research was partly funded by the National Natural Science Foundation of China (Nos. 11972382) and partly funded by KIT the Excellence Strategy of the German Federal and State Governments through the Future Fields Project "Kadi4Mat". Further we thank the Ministry of Science, Research and Art Baden-Württemberg (MWK-BW), in the project MoMaF–Science Data Center, with funds from the state digitization strategy digital@bw (project number 57). The authors acknowledge the support to the research through the "Cluster of Excellence" POLIS of the Deutsche Forschungsgemeinschaft (project number 390874152) and the Helmholtz association within the programme MSE no. 43.31.01. The authors also thank Leon Geisen for editorial support.

\section*{Author contributions statement}
B.N. and Y.Z. supervised and conceived the project. Y.J. and P.A. conducted the multi-phase field simulations and analyzed the results. Y. J. and A.K built the machine learning model, trained the model, and analyzed the results. Y.J. wrote the manuscript, A.K., P.A., L.G., D.R., B.N., W.C., Y.Z. and Y.Z. reviewed and revised the manuscript.

The authors declare no competing interests.

\bibliographystyle{elsarticle-num} 
\bibliography{cas-refs}

\begin{thebibliography}{10}
\expandafter\ifx\csname url\endcsname\relax
  \def\url#1{\texttt{#1}}\fi
\expandafter\ifx\csname urlprefix\endcsname\relax\def\urlprefix{URL }\fi
\expandafter\ifx\csname href\endcsname\relax
  \def\href#1#2{#2} \def\path#1{#1}\fi

\bibitem{caloComputationalComplexityMemory2011}
V.~M. Calo, N.~O. Collier, D.~Pardo, M.~R. Paszyński,
  \href{https://www.sciencedirect.com/science/article/pii/S1877050911002596}{Computational
  complexity and memory usage for multi-frontal direct solvers used in p finite
  element analysis}, Procedia Computer Science 4 (2011) 1854--1861.
\newblock \href {https://doi.org/10.1016/j.procs.2011.04.201}
  {\path{doi:10.1016/j.procs.2011.04.201}}.
\newline\urlprefix\url{https://www.sciencedirect.com/science/article/pii/S1877050911002596}

\bibitem{pivovarovPeriodicBoundaryConditions2019}
D.~Pivovarov, R.~Zabihyan, J.~Mergheim, K.~Willner, P.~Steinmann,
  \href{https://www.sciencedirect.com/science/article/pii/S0045782519304281}{On
  periodic boundary conditions and ergodicity in computational homogenization
  of heterogeneous materials with random microstructure}, Computer Methods in
  Applied Mechanics and Engineering 357 (2019) 112563.
\newblock \href {https://doi.org/10.1016/j.cma.2019.07.032}
  {\path{doi:10.1016/j.cma.2019.07.032}}.
\newline\urlprefix\url{https://www.sciencedirect.com/science/article/pii/S0045782519304281}

\bibitem{jamshidi3DComputationalMethod2023}
F.~Jamshidi, W.~Kunz, P.~Altschuh, T.~Lu, M.~Laqua, A.~August, F.~Löffler,
  M.~Selzer, B.~Nestler,
  \href{https://linkinghub.elsevier.com/retrieve/pii/S2352492823001034}{A {3D}
  computational method for determination of pores per inch ({PPI}) of porous
  structures}, Materials Today Communications 34 (2023) 105413.
\newblock \href {https://doi.org/10.1016/j.mtcomm.2023.105413}
  {\path{doi:10.1016/j.mtcomm.2023.105413}}.
\newline\urlprefix\url{https://linkinghub.elsevier.com/retrieve/pii/S2352492823001034}

\bibitem{richterMote3DOpensourceToolbox2017}
H.~Richter, \href{https://dx.doi.org/10.1088/1361-651X/aa629a}{{Mote3D}: an
  open-source toolbox for modelling periodic random particulate
  microstructures}, Modelling and Simulation in Materials Science and
  Engineering 25~(3) (2017) 035011, publisher: IOP Publishing.
\newblock \href {https://doi.org/10.1088/1361-651X/aa629a}
  {\path{doi:10.1088/1361-651X/aa629a}}.
\newline\urlprefix\url{https://dx.doi.org/10.1088/1361-651X/aa629a}

\bibitem{rutterChargedSurfacesSlabs2021}
M.~J. Rutter,
  \href{https://iopscience.iop.org/article/10.1088/2516-1075/abeda2}{Charged
  surfaces and slabs in periodic boundary conditions}, Electronic Structure
  3~(1) (2021) 015002.
\newblock \href {https://doi.org/10.1088/2516-1075/abeda2}
  {\path{doi:10.1088/2516-1075/abeda2}}.
\newline\urlprefix\url{https://iopscience.iop.org/article/10.1088/2516-1075/abeda2}

\bibitem{nguyenImposingPeriodicBoundary2012}
V.~D. Nguyen, E.~Béchet, C.~Geuzaine, L.~Noels,
  \href{https://www.sciencedirect.com/science/article/pii/S0927025611005866}{Imposing
  periodic boundary condition on arbitrary meshes by polynomial interpolation},
  Computational Materials Science 55 (2012) 390--406.
\newblock \href {https://doi.org/10.1016/j.commatsci.2011.10.017}
  {\path{doi:10.1016/j.commatsci.2011.10.017}}.
\newline\urlprefix\url{https://www.sciencedirect.com/science/article/pii/S0927025611005866}

\bibitem{wangNovelApproachImpose2017}
R.~Wang, L.~Zhang, D.~Hu, C.~Liu, X.~Shen, C.~Cho, B.~Li,
  \href{https://www.sciencedirect.com/science/article/pii/S026382231631710X}{A
  novel approach to impose periodic boundary condition on braided composite
  {RVE} model based on {RPIM}}, Composite Structures 163 (2017) 77--88.
\newblock \href {https://doi.org/10.1016/j.compstruct.2016.12.032}
  {\path{doi:10.1016/j.compstruct.2016.12.032}}.
\newline\urlprefix\url{https://www.sciencedirect.com/science/article/pii/S026382231631710X}

\bibitem{zhouUniversalityDeepConvolutional2020}
D.-X. Zhou,
  \href{https://www.sciencedirect.com/science/article/pii/S1063520318302045}{Universality
  of deep convolutional neural networks}, Applied and Computational Harmonic
  Analysis 48~(2) (2020) 787--794.
\newblock \href {https://doi.org/10.1016/j.acha.2019.06.004}
  {\path{doi:10.1016/j.acha.2019.06.004}}.
\newline\urlprefix\url{https://www.sciencedirect.com/science/article/pii/S1063520318302045}

\bibitem{kalidindiHandbookBigData2020}
S.~Kalidindi, S.~V. Kalinin, T.~Lookman, K.~K.~v. Dam, K.~Yager, S.~Campbell,
  R.~Farnsworth, M.~v. Dam, I.~Foster, Handbook {On} {Big} {Data} {And}
  {Machine} {Learning} {In} {The} {Physical} {Sciences}, World Scientific,
  2020.

\bibitem{nestlerPhasefieldModelMultiphase2008}
B.~Nestler, F.~Wendler, M.~Selzer, B.~Stinner, H.~Garcke,
  \href{https://link.aps.org/doi/10.1103/PhysRevE.78.011604}{Phase-field model
  for multiphase systems with preserved volume fractions}, Physical Review E
  78~(1) (2008) 011604, publisher: American Physical Society.
\newblock \href {https://doi.org/10.1103/PhysRevE.78.011604}
  {\path{doi:10.1103/PhysRevE.78.011604}}.
\newline\urlprefix\url{https://link.aps.org/doi/10.1103/PhysRevE.78.011604}

\bibitem{hotzerParallelMultiphysicsPhasefield2018}
J.~Hötzer, A.~Reiter, H.~Hierl, P.~Steinmetz, M.~Selzer, B.~Nestler,
  \href{http://www.sciencedirect.com/science/article/pii/S1877750317310116}{The
  parallel multi-physics phase-field framework {Pace3D}}, Journal of
  Computational Science 26 (2018) 1--12.
\newblock \href {https://doi.org/10.1016/j.jocs.2018.02.011}
  {\path{doi:10.1016/j.jocs.2018.02.011}}.
\newline\urlprefix\url{http://www.sciencedirect.com/science/article/pii/S1877750317310116}

\bibitem{janssensComputationalMaterialsEngineering2007}
K.~G.~F. Janssens (Ed.), Computational materials engineering: an introduction
  to microstructure evolution, Elsevier / Academic Press, Amsterdam ; Boston,
  2007.

\bibitem{gostickOpenPNMPoreNetwork2016}
J.~Gostick, M.~Aghighi, J.~Hinebaugh, T.~Tranter, M.~A. Hoeh, H.~Day,
  B.~Spellacy, M.~H. Sharqawy, A.~Bazylak, A.~Burns, W.~Lehnert, A.~Putz,
  {OpenPNM}: {A} {Pore} {Network} {Modeling} {Package}, Computing in Science
  Engineering 18~(4) (2016) 60--74, conference Name: Computing in Science
  Engineering.
\newblock \href {https://doi.org/10.1109/MCSE.2016.49}
  {\path{doi:10.1109/MCSE.2016.49}}.

\bibitem{kremeyerCellularAutomataInvestigations1998}
K.~Kremeyer,
  \href{https://linkinghub.elsevier.com/retrieve/pii/S0021999198959265}{Cellular
  {Automata} {Investigations} of {Binary} {Solidification}}, Journal of
  Computational Physics 142~(1) (1998) 243--263.
\newblock \href {https://doi.org/10.1006/jcph.1998.5926}
  {\path{doi:10.1006/jcph.1998.5926}}.
\newline\urlprefix\url{https://linkinghub.elsevier.com/retrieve/pii/S0021999198959265}

\end{thebibliography}





\clearpage

\begin{figure}[ht]
\centering
\includegraphics[width=\textwidth]{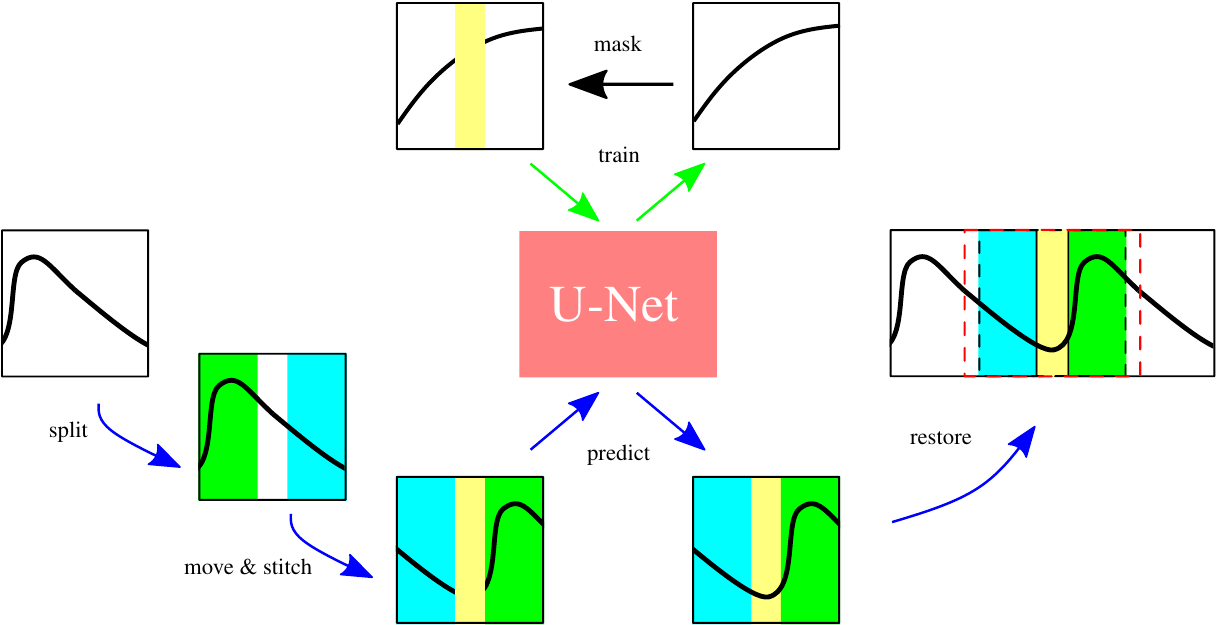}
\caption{Overview of the proposed stitching method. In the first pass (top of the figure), the original data is masked and fed into U-Net for training until U-Net can reconstruct the original data. In the second pass (bottom of the figure), a sample is first split from the middle, and then the two parts are swapped and stitched. A masked area is inserted, and two parts of the mask's half-width are removed at the new boundaries to fit the input size of the U-Net. The removed parts plus the predicted structure is the final periodic structure.}
\label{fig:stitch}
\end{figure}

\begin{figure}[ht]
\centering
\includegraphics[width=\textwidth]{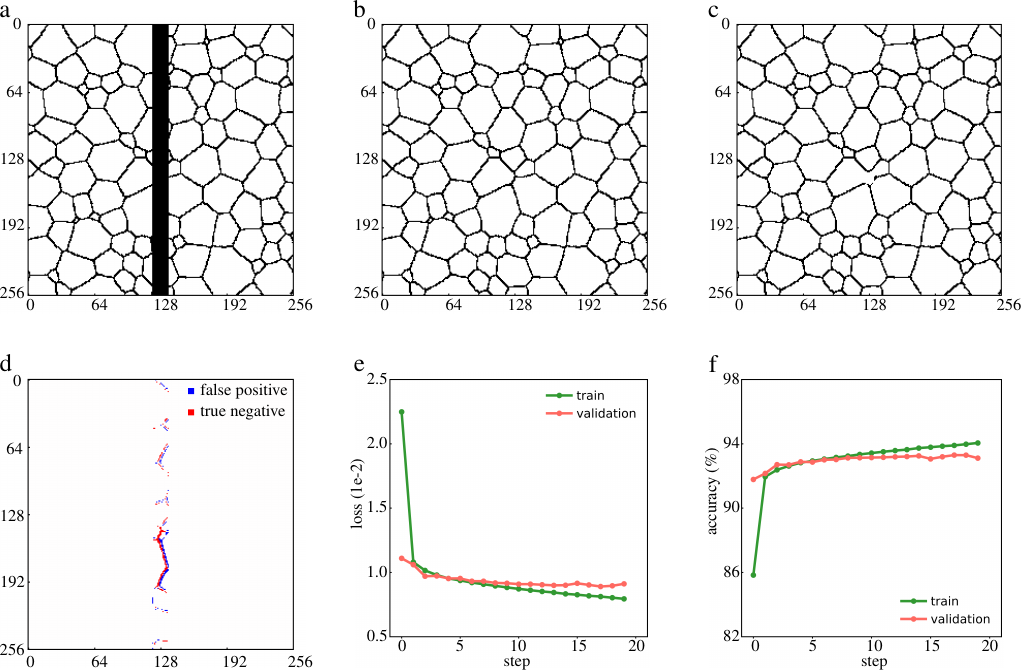}
\caption{The training performance: (a) a grain sample with a 16-cell-width all-zeros masked area in its center along the x-direction throughout the y-direction, (b) the original unmasked  sample as the training target, (c) the reconstructed sample corresponding to the original sample, (d) the differences between the original sample and the reconstructed sample, (e) the training and the validation loss curves, and (f) the training and the validation accuracy curves. }
\label{fig:train}
\end{figure}

\begin{figure}[ht]
\centering
\includegraphics[width=\textwidth]{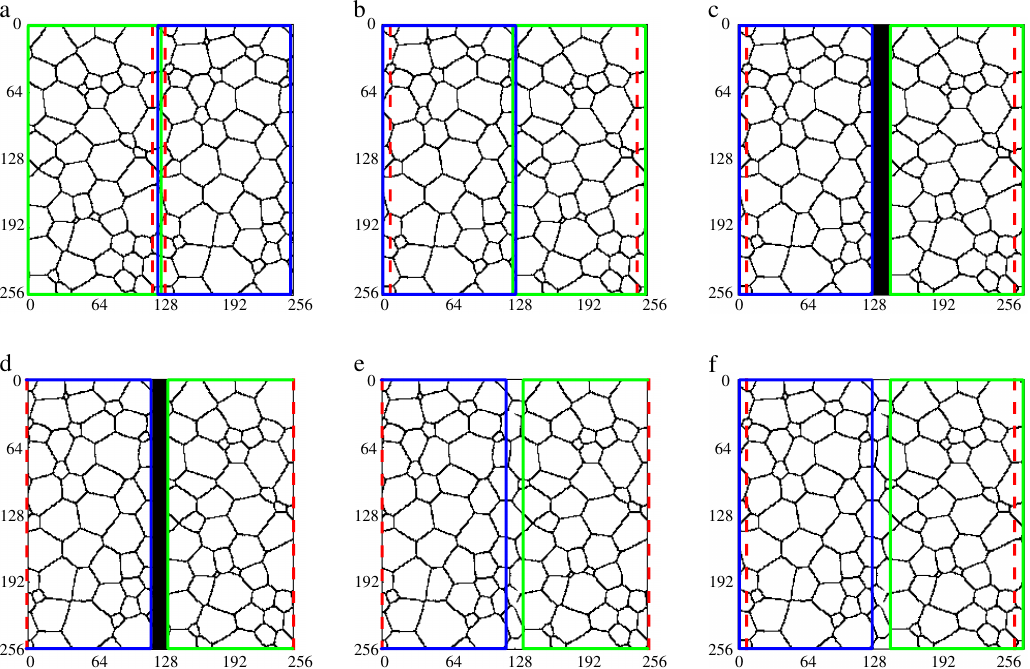}
\caption{The prediction procedure: (a) splitting the original sample into two parts from the middle, (b) switching the left and right parts, (c) inserting an all-zeros area between the two switched parts, (d) cropping the data to fit the input size of U-Net, (e) the prediction from the U-Net, and (f) a periodic sample obtained after restoring the cropped area. For visual guidance, the green and blue frames show the left and right parts in the original sample, respectively. The red dotted line marks the area to be masked.}
\label{fig:predict}
\end{figure}

\begin{figure}[ht]
\centering
\includegraphics[width=\textwidth]{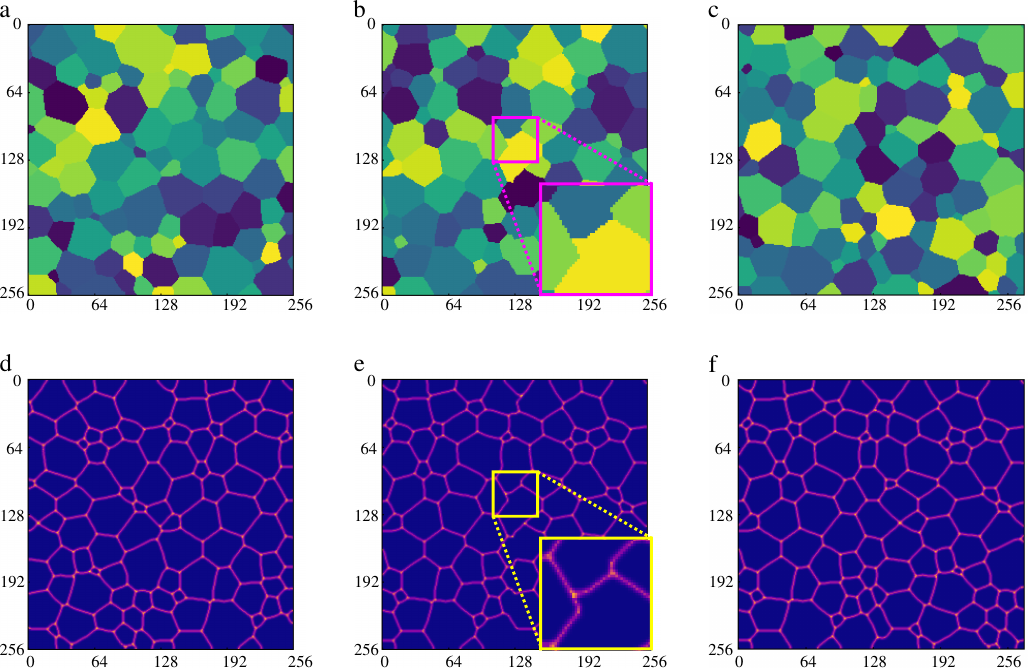}
\caption{The evaluation of the stitching: segmentation of (a) original configuration, (b) switched configuration, and (c) stitched configuration, corresponding to Fig. 3a, 3b, and 3f, respectively. (d)-(f) the grain boundary map illustrating curvature of (a)-(c), respectively.}
\label{fig:evaluate}
\end{figure}

\begin{figure}[ht]
\centering
\includegraphics[width=\textwidth]{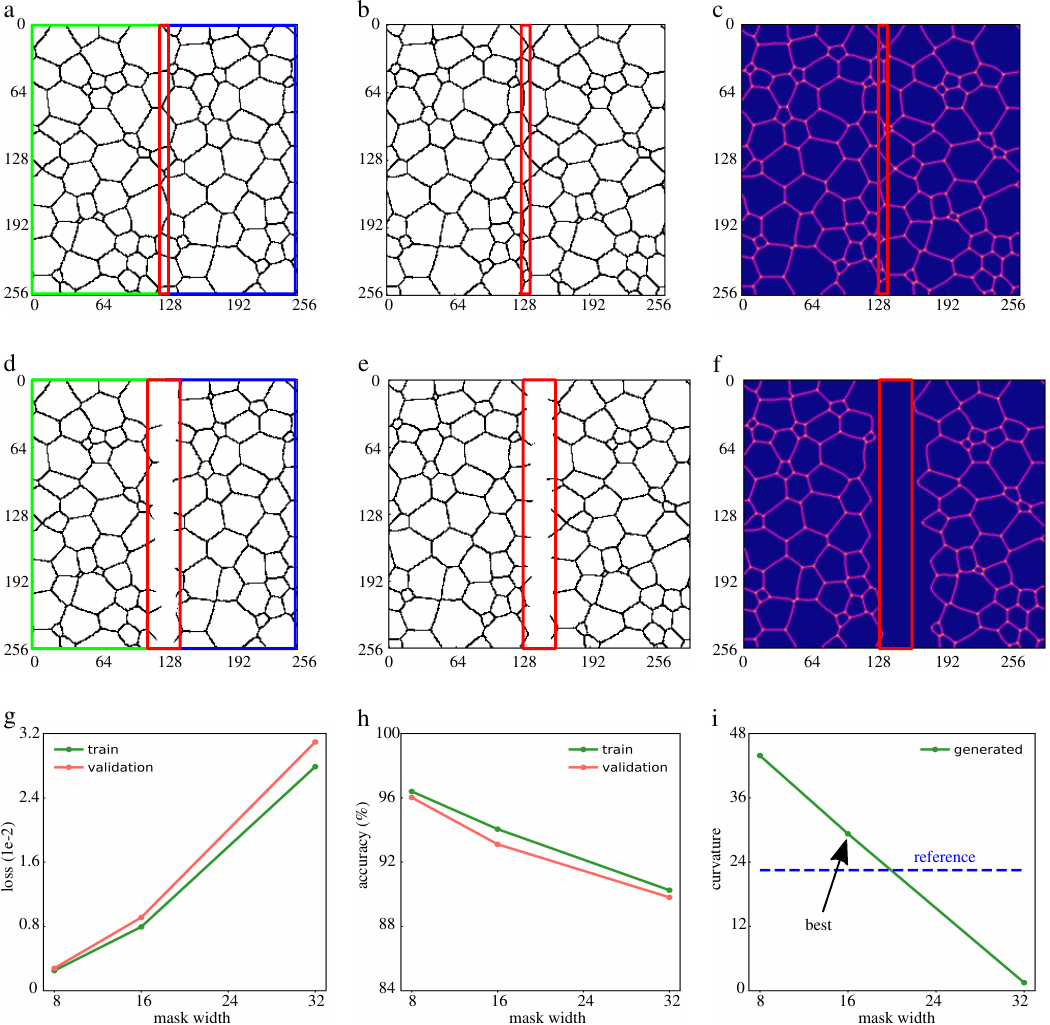}
\caption{The effect of varying mask width: (a) the predicted sample, (b) the stitched sample, (c) the curvature of stitched sample trained with the mask width being 8, (d) the predicted sample, (e) the stitched sample, and (f) the curvature of stitched sample trained with the mask width being 32, comparison of (g) the training loss, (h) the accuracy, and (i) curvature of generated structures under different mask widths. }
\label{fig:optimize}
\end{figure}

\end{document}